\documentstyle[12pt]{article}

\newcommand{\disp}{\displaystyle}
\begin{document}

\begin{center}
{\large\bf Scaling Law for a Magnetic Impurity Model
with Two-Body Hybridization }
\end{center}
\bigskip
\begin{center}
Y. Yu$^*$, Y. M. Li and N. d'Ambrumenil \\
Department of Physics, University of Warwick,\\
Coventry, CV4 7AL, U. K. \\
\end{center}
\bigskip
\bigskip
\begin{center}
{\it ABSTRACT}
\end{center}
{\it We consider a magnetic impurity coupled to the hybridizing and
screening channels of a conduction band.
The model is solved in the framework of poor man's scaling and
Cardy's generalized theories. We point out that it is important to include a
two-body hybridization if the scaling theory  is to be
valid for the band width
larger than $U$. We map out the boundary of the
Fermi-non-Fermi liquid phase transition as a function of the model
parameters.}

\vskip 7cm

\begin{flushleft}
* Address after September 1, 1994: Physics Department, University of Utah,
Salt Lake City, UT84112, USA.
\end{flushleft}
\eject

\bigskip

It has been known for some time that at
energies much less than the Kondo temperature
the ground states of simple models of magnetic impurities $[1,2]$
are Fermi liquids  ${[3,4]}$. Generalized versions of such models show a
transition between Fermi liquid and non-Fermi liquid ground states
[5,6,7,8]. It has recently been argued that analysis of this transition
may lead to a better understanding of the apparently non-Fermi liquid-like
behavior observed in the superconducting cuprates.

Non-Fermi liquid states appear in the phase diagrams of impurity
models with finite range interactions. A spinless impurity model
has been studied by multiplicative renormalization group ${[9]}$
in [6] and using poor man's scaling theory ${[10]}$ in [7].
For a magnetic impurity model, non-Fermi liquid phases have also been found
[5] using Wilson's numerical renormalization group
{[3]}. Something similar also occurs in a generalized Hubbard model
in infinite dimensions ${[8]}$.
In this paper, we adapt poor man's scaling theory and its extension [11]
to find the phase diagram for a magnetic impurity model at finite $U$
including the effect of screening channels
in the particle-hole symmetric case. Until now scaling
arguments have only been developed for the case of infinite $U$.
We point out that for $U$ less than the
bandwidth, a fully renormalizable model must include a `two-body
hybridization process'
if the renormalization group equations are to remain consistent.
This is the process in which two electrons with opposite spin hop from the
conduction band directly onto the local orbital.

For finite $U$ and in the particle-hole symmetric case, we find that
there are four regions
in the phase diagram. These are characterized by three quantities, the usual
hybridization $t_1$, a two-body hybridization $t_2$ and the spin exchange
$V_y$. The first is the non-Fermi liquid fixed point at which all of the
quantities are irrelevant.
The second region corresponds to what we call the
`$V_y$ relevant region', where
the fixed point Hamiltonian belongs to the same universality class of
Kondo's strong fixed point model, and is therefore,
the Fermi liquid fixed point.
The third one is the `$t_2$ relevant region'. We have calculated the exponent
of the leading term of the fixed point interaction Hamiltonian and found
that this fixed point is again the Fermi-liquid fixed point.
We find that the region where all of $g_1$, $g_y$ and $g_2$ are relevant
can be divided into three sections. Two of them
have the same fixed point as that of the $g_y$-relevant and $t_2$-relevant
regions. In the other section the system behaves as `free orbital
Fermi liquid'[2].

Our model Hamiltonian is

\begin{equation}\begin{array}{rcl}\disp
H&=&\disp\sum_{k>0,\sigma,l}\epsilon_kc^\dagger_{k\sigma l} c_{k\sigma l}
+\epsilon_d n_d+Un_{d\uparrow}n_{d\downarrow}\\[3mm]
{}~&+&\disp{t_1}\sum_{\sigma}(c^\dagger_{\sigma 0}d_\sigma+h.c.)
+t_2\sum_\sigma (c^\dagger_{\sigma 0}c^\dagger_{-\sigma 0}d_\sigma
d_{-\sigma}+h.c.)\\[3mm]
{}~&+& \disp\sum_{\sigma,\sigma',l}V_lc^\dagger_{\sigma l}c_{\sigma l}
d^\dagger_{\sigma'}d_{\sigma'}+\sum_{\sigma,l}V_{x l}c^\dagger
_{\sigma l}c_{\sigma l}d^\dagger_\sigma d_\sigma+\sum_{\sigma, l}
V_{y,l}c^\dagger_{\sigma l}c_{-\sigma l}d^\dagger_{-\sigma}d_\sigma.
\end{array}\end{equation}
The first four terms in (1) are the usual Anderson model while $V_l$ term
takes into account the finite range interactions between the local `$d$
orbital' and conduction electrons ($l=0,...,N_f$).
The $V_{x,l}$ and $V_{y,l}$
terms describe the spin exchange interaction and are taken to be zero for
$l=1,...,N_f$ for convenience.
The $t_2$ term
${[12]}$ describes processes in which two $l=0$-channel conduction
electrons with opposite spin
hop onto the local orbital.
To renormalize the model in the framework of poor man's scaling theory
it is important that this term as well as spin exchange terms are included.

Following the terminology of [5,6], we call the 0-channel
the hybridizing channel
and the others the screening channels. We deal with the screening channels
by bosonization. The spin of the screening channels is not important and we
treat the screening channel as spinless for simplicity.
Let $b_{k l}$ be the bosonic operators corresponding
to $c_{k l}$ for $l=1,...,N_f$ [13]. The effective Hamiltonian then is
$H=H_0+H_I$, where

\begin{equation}\begin{array}{rcl}\disp
H_0&=&\disp\sum_{k>0,\sigma}\epsilon_kc^\dagger_{k\sigma} c_{k\sigma}
+\epsilon_dn_d+U~n_{d\uparrow}n_{d\downarrow}\\[3mm]
{}~&+&\disp V_0\sum_{\sigma,\sigma'}c^\dagger_\sigma c_\sigma
d^\dagger_{\sigma'}d_{\sigma'}+V_x\sum_\sigma c^\dagger_\sigma
c_\sigma d^\dagger_\sigma d_\sigma +\sum_{k,l>0,\sigma}\frac{k}{\rho}
b^\dagger_{kl}b_{kl},\\[5mm]
H_I&=&\disp{t_1}\sum_\sigma(\Delta^\dagger
c^\dagger_\sigma d_\sigma
+h.c.)+t_2\sum_{\sigma}(\Delta^{\dagger 2}c^\dagger_\sigma
c^\dagger_{-\sigma}d_\sigma d_{-\sigma}+h.c.)+V_y\sum_\sigma
c^\dagger_\sigma c_{-\sigma}d^\dagger_{-\sigma}d_\sigma,
\end{array}\end{equation}
where the index 0 denoting the $l=0$ channel has been suppressed.
The electron dispersion is taken to be $\epsilon_k=(k-k_F)/\rho$ with $\rho
=(hv_F)^{-1}$ being the density of states at the Fermi surface. The operator
$\Delta$ is given by $\Delta=\exp\{-\sum_{k}\rho V_l/\sqrt{kL}(b_{k l}-
b^\dagger_{kl})\}$.
The Hilbert space of $H_0$  can be projected onto four subspaces
characterized by the four possible impurity electron
states, $|\alpha>=|0>, |\sigma>$ and
$|3>\equiv|\uparrow\downarrow>$. Each term in the
interaction $H_I$ plays the role
of a dipole operator causing transition among those subspaces.

The Hamiltonian in (2) can be treated using Haldane's familiar procedure
${[14]}$ by writing the partition function
in terms of a sum over histories of the impurity.
Expanding the partition function in $H_I$ and labelling
the (imaginary) times so that
$0<\tau_1<...<\tau_n<\beta$, the partition function
can eventually be written as a sum over all possible histories of the
local degrees of freedom which fluctuate between the
4 local states. This gives
\begin{equation}\begin{array}{l}
Z=\disp\sum_{n}\sum_{\alpha_1,...,\alpha_n}
\int_0^{\beta-\tau}\frac{d\tau_n}{\tau}...\int_0^{\tau_2-\tau}\frac
{d\tau_1}{\tau}
\exp(-S[\tau_1,...,\tau_n;\alpha_1,...,\alpha_n]),\\[6mm]
S[\tau, \alpha]=\disp\sum_{i<j}\sum_{a=\sigma,l}q^a_{\alpha_i\alpha_{i+1}}
q^a_{\alpha_j\alpha_{j+1}}\ln\frac{\tau_j-\tau_i}{\tau}-\sum_i\ln(g_{\alpha_i
\alpha_{i+1}})+\sum_iE_{\alpha_{i+1}}\frac{\tau_{i+1}-\tau_i}{\tau},
\end{array}\end{equation}
where $\alpha_1,...,\alpha_n$ ($\alpha_{n+1}=\alpha_1$)
and $\tau_1,...,\tau_n$
label a Feynman trajectory. $\tau$ is the ultraviolet cut-off.
With a simple shift of the ground state energy, $E_\alpha$
can be chosen so that
$\sum_\alpha E_\alpha=0$. We take $E_0=E_3=-E_\sigma=-\epsilon_d\tau
/2$ for the particale-hole symmetric case ($\epsilon_d=-U/2$).
$g_{\alpha\beta}$ are coupling constants,
with $g_{0\sigma}=g_{\sigma 3}\equiv g_1=t_1\tau$,
$g_{\sigma,-\sigma}
\equiv g_y=V_y\tau$ and $g_{03}\equiv g_2=t_2\tau$.

Equation (3) may be thought of as describing
a one-dimensional four-component plasma of kinks carrying
`charges' $q^a_{\alpha
\beta}$ and `fugacities' $g_{\alpha\beta}$
[11,8]. $E_\alpha$ is a `magnetic field'.
The `charges' are given by $q^a_{0\sigma}=((1-\delta_x/\pi)\delta_
{\lambda\sigma}-\delta_0/\pi,\delta_l/\pi)$, $q^a_{\sigma,\sigma'}=
((1-\delta_x/\pi)(\delta_{\lambda\sigma'}-\delta_{\lambda\sigma}),0)$,
$q^a_{\sigma 3}=((1-\delta_x/\pi)\delta_{\lambda,-\sigma}
-\delta_0/\pi,\delta_l/\pi)$ and $q^a_{03}=((1-\delta_x/\pi)-2\delta_0/\pi,
2\delta_l/\pi)$. $q_{\alpha\beta}=-q_{\beta\alpha}$. The phase shifts
are
$\delta_0=2\tan^{-1}\pi\rho V_0/2$, $\delta_x=2\tan^{-1}\pi\rho V_x/2$ and
$\delta_l=2\tan^{-1}\pi\rho V_l/2$ [17].
The `charges' obey the relations
$q_{\alpha\beta}+q_{\beta\gamma}=q_{\alpha\gamma}$, which means that the
model can be regarded as a special case of the general one-dimensional model
with $1/r^2$ interaction considered by Cardy ${[11]}$ and can be
renormalized by poor man's scaling theory. The Coulomb term in (3)
can be rewritten as a one-dimensional spin chain model with interaction
$\disp\sum_{i<j}K(\alpha_i,\alpha_j)\tau^2/(\tau_i-\tau_j)^2$, where
$K(\alpha,\beta)=-1/2\sum_a(q^a_{\alpha\beta})^2$, {\it i.e.}
$K(0,\sigma)=K(\sigma,3)=-\gamma_0$, $K(\sigma,\sigma')=-\gamma_x(1-\delta_{
\sigma,\sigma'})$ and $K(0,3)=\gamma_x-2\gamma_0$ with
$K(\alpha,\beta)=K(\beta,\alpha)$ and $K(\alpha,\alpha)=0$. Here
$\gamma_0=(1-\delta_x/\pi-\delta_0/\pi)^2+(\delta_0/\pi)^2
+\sum_l(\delta_l/\pi)^2$ and $\gamma_x=(1-\delta_x/\pi)^2$. It is worth
noting that if we consider the model with vanishing $t_2$, the relations
$q_{\alpha\beta}+q_{\beta\gamma}=q_{\alpha\gamma}$ are not satisfied so that
the model can not be mapped to the Cardy's model.

Cardy has pointed out that the critical behavior of these
kinds of model can be
discussed using
a generalized poor man's scaling theory. Si and Kotliar have
extended Cardy's theory to the case in which small `magnetic fields'
$E_\alpha$ are included ${[15]}$
as Haldane has also done ${[14]}$.
For the particle-hole symmetric case of the present model,
the renormalization group
equations are given by
\begin{equation}\begin{array}{l}\disp
\frac{dg_1}{d\ln\tau}=\disp
(1-\frac{1}{2}\gamma_0)g_1-g_1g_ye^{-E_\sigma}-g_1g_2
e^{-E_0},\\[3mm]\disp
\frac{dg_y}{d\ln\tau}=\disp(1-\gamma_x)g_y-2g_1^2e^{-E_0} ,\\[3mm]
\disp\frac{dg_2}{d\ln\tau}=\disp(1-2\gamma_0+\gamma_x)g_2-2g_1^2e^{-E_

\sigma},\\[3mm] \disp
\frac{d\gamma_0}{d\ln\tau}=\disp-8\gamma_0(g_1^2e^{-E_\sigma}+g_2^2e^{-
E_0})+4\gamma_x[g_1^2(e^{-E_\sigma}-e^{-E_0})+g_2^2e^{-E_0}-g_y^2e^{-E_
\sigma}],\\[3mm]\disp
\frac{d\gamma_x}{d\ln\tau}=\disp-4\gamma_x(g_1^2e^{-E_0}+g_y^2e^{-E_\sigma})
,\\[3mm]\disp
\frac{d\epsilon_d\tau}{d\ln\tau}=\disp\epsilon_d\tau-2g_1^2(e^{-E_0}-
e^{-E_\sigma})+2g_2^2e^{-E_0}-2g_y^2e^{-E_\sigma}.
\end{array}\end{equation}

We make some comments on these equations.
The first three equations of (4) are exact for any value of
$\epsilon_d\tau$
 while the last three are restricted to the
case where the impurity level $\epsilon_d$
is much less than the band width $1/\tau$ ${[14]}$. ( The assumption
of a large band width
is within  the spirit of poor man's scaling theory.) The last equation
reflects the variation of $\epsilon _d$ with $1/\tau$:
\begin{equation}
\tau\frac{d\epsilon_d}{d~\ln\tau}=-2g_1^2(e^{-E_0}-
e^{-E_\sigma})+2g_2^2e^{-E_0}-2g_y^2e^{-E_\sigma}.
\end{equation}
The requirement $\sum_\alpha E_\alpha=0$ which we enforce throughout the
scaling process, ensures that the separation of the field $\epsilon_d\tau$
renormalization and the free energy renormalization is unambiguously
determined.

We have assumed that $g_{0\sigma}=g_{\sigma 3}$,
$K(0,\sigma)=K(-\sigma,3)$ and
$K(0,3)=4K(0,\sigma)-K(\sigma,-\sigma)$ in the present model. The
scaling must preserve these equalities if the model is to remain consistent.
This only happens in the particle-hole symmetric case.
Away from the particle-hole symmetry, the scaling
generates different flows for $g_{0\sigma}$ and $g_{\sigma 3}$,
$K(0,\sigma)$ and $K(-\sigma,3)$, {\it etc}.

As we have already said, if the model can be mapped to the special case
of Cardy's model, $t_2$ can not vanish. Furthermore, from (4), we
can see that even if we start with vanishing $g_2$ and $g_y$,
their absolute values all increase at a rate proportional to
$g_1$. This means that the
flows calculated assuming that $g_y$ and $ g_2$
vanish exactly are not correct renormalization
flows. To describe a fully renormalizable
model when the band width is larger than $U$, the Hamiltonian
must include these two-body hybridization and spin exchange terms.

Since our renormalization group is perturbative in its treatment of
$g_{\alpha\beta}$, the renormalization of $\gamma_0$ and $\gamma_x$
can be neglected in the first three equations of (4).
This allows us to draw out the phase diagram in
$\gamma_0-\gamma_x$ space (See Figure 1), which can be divided into
four regions:

(i) For $\gamma_0>2$, $\gamma_x>1$ and $2\gamma_0-\gamma_x>1$, all
$g_{\alpha\beta}$ are irrelevant and renormalize to zero. There exist
weak coupling fixed points $g^*_{\alpha\beta}=0$. The fixed
point Hamiltonian is similar to the multi-channel X-ray edge problem ${[16]}$.
The system exhibits a power-law decay of the correlation function with
a non-universal exponent. This is the non-Fermi liquid phase.
Since $\gamma_0=(1-\delta_0/\pi-\delta_x/\pi)^2+(\delta_0/\pi)^2+\sum_l
(\delta_l/\pi)^2$, it is clear that it is the existence of the screening
channels which allows $\gamma_0$ to exceed 2 and leads to the non-Fermi liquid
phase [7].

(ii) For $\gamma_0>2$, $2\gamma_0-\gamma_x>1$ but $\gamma_x<1$, $g_1$ and
$g_2$ are irrelevant and renormalize to zero. $g_y$ is relevant.
We focus on the fixed point Hamiltonian. Choosing the parameters
$V_y=J_\perp$,
$V_0=2J_z$ and $V_x=-J_z$ with $J_z<0$ (this ensures $\gamma_x<1$),
we see that the fixed point
Hamiltonian $H^*$ is just the single-channel Kondo Hamiltonian:
\begin{equation}\begin{array}{rcl}\disp
H^*&=&\disp\sum_{k>0,\sigma}\epsilon_k
c^\dagger_{k\sigma}c_{k\sigma}+\epsilon_d n_d
+Un_{d\uparrow}n_{d\downarrow}+\sum_{k>0,l=1}\frac{k}{\rho}b^\dagger_{kl}
b_{kl}\\[3mm]\disp
&+&\disp\frac{J_\perp}{2}(S^+_c s^-_d+h.c.)+J_zS^z_cs^z_d.
\end{array}\end{equation}
Here ${\bf S}_c$ and ${\bf s}_d$ are the spin operators of hybridizing
and localized electrons respectively. This $g_y$ ($J_\perp$)-relevant
region is controlled by the strong-coupling fixed point of the Kondo problem.
Furthermore, the values of the interaction parameters $V_0$ and
$V_x$ are not important as long as they are within the parameter regime of
the $g_y$ relevant region. Hence, the fixed point corresponding to this
$g_y$ relevant phase is the Fermi liquid fixed point. Near
the fixed point equation (5) reduces
to, $\disp\tau\frac{d\epsilon_d}{d~\ln\tau}=-2g_y^2$, which
implies that $U$ increases. This is consistent with the results
for symmetric Anderson model: when the impurity level width $\Gamma\sim
t_1^2$ is much less than $U$, the model is equivalent to the Kondo model
[2,18]. We therefore call the $g_y$-region the Kondo region.

(iii) For $\gamma_0>2$, $\gamma_x>3$ but $2\gamma_0-\gamma_x<1$, $g_1$ and
$g_y$ are irrelevant and $g_2$ is
relevant. This parameter regime corresponds to
the empty and doubly-occupied states being
favored over singly-occupied states. We call this the $G_2$-region.
The fixed point Hamiltonian $H^*=H^*_0+H^*_I$ where
$H^*_0$ can be read off from (2) by replacing  all  parameters by
their fixed point values. $H^*_I$ has its leading term $R
=(g^*_2/\tau)\sum_{\sigma}
(c^\dagger_\sigma c^\dagger_{-\sigma}d_\sigma d_{-\sigma}+h.c.)$.
The matrix elements of $R$ in the fixed point basis
can be evaluated by analogy with the X-ray
edge problem ${[19]}$. As was done in [5], we denote by $|n_d=0>$ the
eigenstates of $H^*_0|_{n_d=0}$ and by $|n_d=2>$ the
eigenstates of $H^*|_{n_d=2}$.
Then
\begin{equation}
<n_d=0|R|n_d=2>\sim \tau^{\alpha}.
\end{equation}
The anomalous exponent $\alpha= (1-2\gamma_0+\gamma_x)/2$.
If $\alpha<0$ the operator $R$ is irrelevant, and the fixed
point corresponds to the first kind
of fixed point we have discussed.
In the present parameter regime, $\alpha>0$ and
the operator $R$ is relevant. The fixed point is
regarded as a Fermi liquid one ${[5]}$.
The line $\alpha=0$ is marginal.
(We note in passing that if $V_x=0$, and hence $\gamma_x=1$, our result
 appears to recover that of [5]. However, as the parameter region in which
(7) holds requires $\gamma_x>3$, this may be just a coincidence.)

(iv) For $\gamma_0<2$,
all coupling constants are relevant. We divide this parameter region into
three sections (Figure 1).
Although all coupling constants are relevant, (4) shows that
$g_y^2\gg g_2^2$ in $I$, $g_2^2\gg g_y^2$ in $II$ and $g_y^2\sim g_2^2$
in $III$. According to (5), the Hubbard interaction $U$ increases
and so singly-occupied states are favored in $I$ which is in the
Kondo strongly coupling phase. In $II$,  $U$ decreases and  the
empty and doubly-occupied states are favored.
This suggests that in $II$ the system
is the same FL phase as the $G_2$-region. In $III$,
$U$ remains close to its initial small value and  $|0>$, $|\sigma>$ and
$|3>$ are mixed. The impurity level width
$\Gamma(\sim t_1^2)\gg U$ in $III$ implies that the system
is in the `free orbital Fermi liquid phase' [2].

We may summarise what we have learnt about the phase diagram as follows.
There are two kinds of fixed point in the phase diagram: Fermi liquid and
non-Fermi liquid. The part of the phase diagram controlled by
Fermi liquid fixed point can be divided into three regions according to the
behaviour expected at finite temperature. The three types of behaviour are
Kondo strong-coupling, free-obital and what we have called `$G_2$'.

It is interesting to compare our results with those of the numerical
renormalization group reported
by Krishna-murthy, Wilson and Wilkins [2]. The case
they studied corresponds to a
point in our phase diagram: $\gamma_0=\gamma_x=1$.
For this case, we reproduce qualitatively their results from the
renormalization group equations (4). They found
that the system goes to a Fermi liquid
fixed point with two regions: which they call free orbital
region (this corresponds to our region $III$
in Fig. 1) and local moment region ($I$ in Fig. 1).
Because they assume that $t_2$ is always zero, it is clear that they
would not find the region corresponding to our region $II$.  The `$G_2$-'
and `NFL-' regions in Fig. 1 are
physically relevant. If there is a direct Coulomb interaction
between the conduction electrons and the impurity electron ({\it i.e.
} $V_0\not=0$
and $V_x\not=0$), the system could be driven into the $II$-region
even if the two-body hybridization vanishes initially.

In conclusion, we have derived that the scaling laws for a magnetic impurity
model including the `hybridization' of
an up and a down spin
electron hopping onto the local orbital.
The renormalization group analysis shows that
there is a Fermi-non-Fermi liquid transition when the
strength of the local interaction is varied.
For the case of finite $U$ which we have considered, the
model has to be particle-hole
symmetric in order to preserve
the consistency of the renormalization group equations derived from
the generalized poor man's scaling theory.

The asymmetric model
can be discussed only in the infinite $U$ limit, when the
doubly-occupied states are completely suppressed leaving only local states
$|0>$ and $|\sigma>$. This is then just a special case of the spin-$N+1$ model
with added screening channels and has been discussed by Si and Kotliar [15].
Results of a numerical renormalization group calculation were also reported in
[5]. At finite $U$, the normal assumption that the usual hybridization remains
particle-hole symmetric after renormalization is valid only for
the particle-hole symmetric case. Results of
more general treatment will published separately ${[20]}$.

The authors thank Professor Z. B. Su who brings us to pay our
attention to this type of problems and presented a prior.
We are also very grateful for useful discussions
with him and Professor L. Yu. This work was supported
in part by SERC of the United Kingdom under grant No.GR/E/79798 and also by
MURST/British Council under grant No.Rom/889/92/47.

\eject

\begin{description}

\item{[1]} J. Friedel, Can. J. Phys. {\bf 34}, 1190 (1956);
A. Blandin and J. Friedel, J. Phys. Radium {\bf 20}, 160 (1959).
\item{[2]} P. W. Anderson, Phys. Rev. {\bf 124}, 41(1961); H. R.
Krishna-murthy, K. G. Wilson and J. W. Wilkins,
Phys. Rev. Lett. {\bf 16}, 1101 (1975).
\item{[3]} K.G. Wilson, Rev. Mod. Phys. {\bf 47}, 773(1975).
\item{[4]} P. Nozieres, J. Low Temp. Phys. {\bf 17}, 31(1974).
\item{[5]} I. E. Perakis, C. M. Varma and A. E. Ruckenstein,
Phys. Rev. Lett. {\bf 70}, 3467 (1993).
\item{[6]} T. Giamarchi, C. M. Varma,
A. E. Ruckenstein and P. Nozieres, Phys. Rev. Lett. {\bf 70},
3967 (1993).
\item{[7]} G. M. Zhang, L. Yu and Z. B. Su, Phys.Rev. {\bf B49}, 7759 (1994).
\item{[8]} Q. M. Si and G. Kotliar, Phys. Rev. Lett. {\bf 70},
3143 (1993).
\item{[9]} J. Solyom, Adv. Phys. {\bf 28}, 209 (1979); J. Phys.
{\bf F 4}, 2269 (1975); P. Nozieres,and A. Blandin, J. Phys.
(Paris) {\bf 41}, 193 (1980).
\item{[10]} P. W. Anderson, G. Yuval and D. R. Hamann, Phys. Rev.
{\bf B1}, 4464 (1970); P. W. Anderson, J. Phys. {\bf C3}, 2436.
\item{[11]} J. L. Cardy, J. Phys. {\bf A14}, 1407 (1981); See also
S. Chakravarty and J. Hirsch, Phys. Rev. {\bf B25}, 3273 (1982) and
Q. Si and G. Kotliar, Phys. Rev. {\bf B48}, 13881 (1993).
\item{[12]} In [5], this term was discussed as the leading term of the
fixed point Hamiltonian.
\item{[13]} D. C. Mattis and E. Lieb, J. Math. Phys. {\bf 6}, 304 (1965).
\item{[14]}F. M. D. Haldane, Phys. Rev. Lett. {\bf 40}, 416 (1978);
J. Phys. {\bf C 11}, 5015 (1978).
\item{[15]} See Q. M. Si and G. Kotliar in refs.[11] and [8].
\item{[16]} G. Mahan, Phys. Rev. {\bf 153}, 882 (1967); P. Nozieres and
C. T. de Domincis, Phys. Rev. {\bf 178}, 1097 (1969); P. W. Anderson,
Phys. Rev. Lett. {\bf 18}, 1049 (1967).
\item{[17]} P. B. Wiegmann and A. M. Feinkel'stein, Sov. Phys.
JEPT {\bf 48}, 102 (1978).
\item{[18]} P. W. Anderson and A. M. Clogston, Bull. Am. Phys. Soc.
{\bf 6}, 124(1961); J. Kondo, Prog. Theor. Phys. {\bf 28}, 846(1962);
J. R. Schrieffer and P. A. Wolff, Phys. Rev. {\bf 149}, 491(1966).
\item{[19]} See, for example, P. Nozieres and C. T. de Dominicis in [16];
Also see K. D. Schotte and U. Schotte, Phys. Rev. {\bf 182}, 479 (1969);
Recent application, see [5].
\item{[20]} Y. Yu, Y. M. Li and N. d'Ambrumenil, in preparation.

\end{description}
\eject
\begin{description}
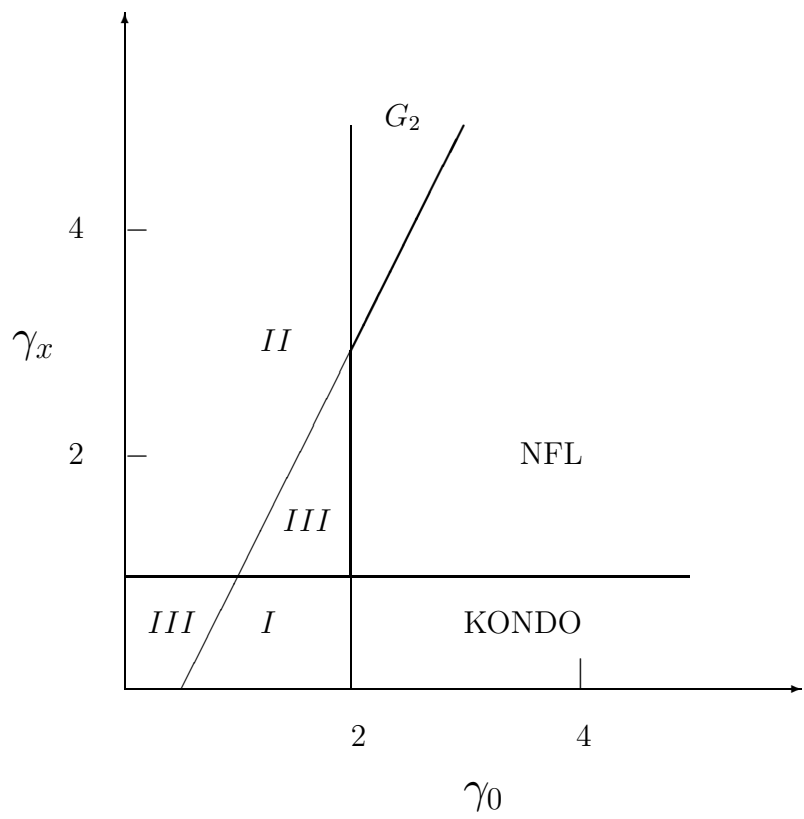
\item{Fig.1} The phase diagram in $\gamma_0-\gamma_x$ space. The thicklines
are phase boundaries.
The thin lines divide the Fermi liquid phase into the different regions
characterised by different behaviour at finite temperature..
\end{description}

\eject
\begin{picture}(8,8)(-60,450)
\unitlength 1.5cm
\put(0.0,0.0){\vector(0,1){6.0}}
\put(0.0,0.0){\vector(1,0){6.0}}
\put(2.0, 0.0){\line(0,1){1.0}}
\put(2.0,3.0){\line(0,1){2.0}}
\put(0.0,1.0){\line(1,0){5.0}}
\put(2.0,1.0){\line(0,1){2.0}}
\put(0.5,0.0){\line(1,2){2.2}}
\put(1.2,0.5){\shortstack{$I$}}
\put(1.2,3.0){\shortstack{$II$}}
\put(1.4,1.4){\shortstack{$III$}}
\put(0.2,.5){\shortstack{$III$}}
\put(-0.5,2.0){\shortstack{2}}
\put(-0.5,4.0){\shortstack{4}}
\put(2.0,-.5){\shortstack{2}}
\put(4.0,-.5){\shortstack{4}}
\put(3.0,-1.0){\shortstack{\Large$\gamma_0$}}
\put(-1.0,3.0){\shortstack{\Large$\gamma_x$}}
\put(0.0,2.0){\shortstack{$-$}}
\put(0.0,4.0){\shortstack{$-$}}
\put(4.0,.06){\shortstack{$|$}}
\put(3.0,.5){\shortstack{KONDO}}
\put(3.5,2.0){\shortstack{NFL}}
\put(2.3, 5.0){\shortstack{$G_2$}}
\put(2.5,-2.0){\shortstack{FIGURE  1}}
\thicklines
\put(2.0, 1.0){\line(1,0){3.0}}
\put(2.0,1.0){\line(0,1){2.0}}
\put(2.0,3.0){\line(1,2){1.0}}
\end{picture}
\end{document}